\documentclass[pra,twocolumn,showpacs,preprintnumbers,superscriptaddress]
{revtex4}

\usepackage{times}
\usepackage{bm}
\usepackage{float}
\usepackage{graphicx}
\usepackage{amsbsy}
\usepackage{amssymb}
\usepackage{amsmath}
\usepackage{amsfonts}
\usepackage{amsthm}
\usepackage{tikz}
\usepackage{caption}
\usepackage{refstyle}
\usetikzlibrary{shapes,arrows, positioning,arrows, automata, quotes}
\begin{document}

\theoremstyle{plain}
\newtheorem{theorem}{Theorem}
\newtheorem{lemma}[theorem]{Lemma}
\newtheorem{corollary}[theorem]{Corollary}
\newtheorem{proposition}[theorem]{Proposition}
\newtheorem{conjecture}[theorem]{Conjecture}

\theoremstyle{definition}
\newtheorem{definition}[theorem]{Definition}

\theoremstyle{remark}
\newtheorem*{remark}{Remark}
\newtheorem{example}{Example}
\title{Detection of $d_{1}\otimes d_{2}$ Dimensional Bipartite Entangled State: A Graph Theoretical Approach}
\author{Rohit Kumar, Satyabrata Adhikari}
\email{rohitkumar@dtu.ac.in, satyabrata@dtu.ac.in} \affiliation{Delhi Technological
University, Delhi-110042, Delhi, India}

\begin{abstract}
Braunstein et. al. have started the study of entanglement properties of the quantum states through graph theoretical approach. Their idea was to start from a simple unweighted graph $G$ and then they have defined the quantum state from the Laplacian of the graph $G$. A lot of research had already been done using the similar idea. We ask here the opposite one i.e can we generate a graph from the density matrix? To investigate this question, we have constructed a unital map $\phi$ such that $\phi(\rho)=L_{\rho}+\rho$, where the quantum state is described by the density operator $\rho$. The entries of $L_{\rho}$ depends on the entries of the quantum state $\rho$ and the entries are taken in such a way that $L_{\rho}$ satisfies all the properties of the Laplacian. This make possible to design a simple connected weighted graph from the Laplacian $L_{\rho}$. We show that the constructed unital map $\phi$ characterize the quantum state with respect to its purity by showing that if the determinant of the matrix $\phi(\rho)-I$ is positive then the quantum state $\rho$ represent a mixed state. Moreover, we study the positive partial transpose (PPT) criterion in terms of the spectrum of the density matrix under investigation and the spectrum of the Laplacian associated with the given density matrix. Furthermore, we derive the inequality between the minimum eigenvalue of the density matrix and the weight of the edges of the connected subgraph of a simple weighted graph to detect the entanglement of $d_{1} \otimes d_{2}$ dimensional bipartite quantum states. Lastly, we have illustrated our results with few examples. 
\end{abstract}
 \maketitle

\section{Introduction}
\noindent Entanglement \cite{horodecki1} is one of the key topic in quantum information theory that lies at the heart of quantum mechanics. This feature of quantum mechanics has no classical analogue and play a vital role in enhancing the power of quantum computation \cite{jozsa}. It also acts as a useful resource in many quantum information processing tasks such as quantum teleportation \cite{bennett}, quantum superdense coding \cite{bennett1}, quantum cryptography \cite{ekert,ngisin} etc. Thus, it is important to generate the entangled state in the laboratory for processing the quantum information tasks. But the experimentalist may face the problem in identifying the generated state: whether the state at the output of the experiment is entangled or not? This type of problem is known as entanglement detection problem.\\
It is one of the prime task for the researcher to develop the criterion for the detection of entanglement. A lot of research have already been carried out in this direction and as a consequence different criterions such as partial transposition criterion \cite{peres,horodecki2}, computable cross norm or realignment criterion \cite{rudolph,chen}, reduction criterion \cite{horodecki3} have been emerged to detect the entangled state. Other criterion for the detection of entanglement can be found in \cite{guhne}.\\
There exist another approach to study the problems in quantum information theory - the graph theoretical approach. Along this line of research,  Braunstein et.al. \cite{braunstein1} have started with any graph $G$ and then associate a specific mixed quantum state with it. Mixed state is described by the density matrix and therefore they called it as the density matrix of graph $G$. In this way, they have introduced the concept of density matrices of graph to study the graphical representation of quantum states and their properties. The entanglement properties of the mixed density matrices obtained from the combinatorial Laplacians has been studied in \cite{braunstein2}. A. Cabello et.al. \cite{cabello} used graph to characterize the correlations with respect to different sets of probabilities obtained through non-contextual theories, quantum theory and more general probabilistic theories. M. Ray et.al. \cite{ray} have taken graph theoretical approach to find the minimum quantum dimension required for performing given quantum task by identifying quantum dimension witnesses. The graphical characterization of the entanglement properties of the grid states has been studied in \cite{lockhart1}. The separability problem of bipartite quantum states generating from graphs has been studied in \cite{dutta}.\\
In this work, we will study the entanglement detection problem using the graph theoretical approach. But our approach is different from the other graph-theoretical approaches that are existing in the literature. In the existing graph theoretical approaches, the density matrices are generated from the Laplacian of the given graph but in our approach, we have constructed an unital map that take the density matrices as the input and at the output, it provides the input density matrix together with other matrix $L$. Later, we prove that the matrix $L$ satisfies all the properties of Laplacian. Then we can construct a graph corresponding to each  Laplacian at the output of the map. Further, we will show that the entanglement property of the quantum states can be studied using the eigenvalues of the generated Laplacian.\\
The work can be distributed in different sections in the following way: In section-II, we discuss about the basics of graph theory and its terminology. In particular, we will talk about a simple weighted graph that may be associated with the quantum state described by the density matrix. In section-III, we describe briefly the tensor formalism in quantum information theory. In section-IV, we revisit few results that have been obtained earlier in the literature. In section-V, we construct a unital map and study the properties of the map. Also, we have discussed the physical interpretation of the total degree of the graph corresponding to the Laplacian generated at the output of the map. In section-VI, we derive the necessary condition to test whether a given quantum state described by the density operator $\rho$ is either a pure state or mixed state. In section-VII, we study the PPT criterion in terms of the eigenvalues of the given $d_{1} \otimes d_{2}$ dimensional bipartite state and the corresponding Laplacian. In section-VIII, we study the entanglement properties using the graph terminology.
\section{Basics of graph theory and the associated terminology}
\noindent Let $G=(V,E)$ be a simple weighted graph (may or may not be connected) with vertex set $V=\{v_{1},v_{2},.....,v_{n}\}$ and edge set $E$. If the two vertices $v_{i}$ and $v_{j}$ are connected by an edge $e_{ij}$ then we write $i\sim j$. A graph is said to be a simple graph if it has no loops and multiple edges. If we assigned a weight $w_{i,j}$ to each edge $e_{i,j}$ in a graph $G$ then the graph is known as a weighted graph. The weight $w_{i,j}$ associated to each edge is usually a positive number. If $w_{i}$ denote the weight of the vertex $v_{i}$ then it can be defined as $w_{i}=\sum_{k \sim i}w_{k,i}$. If $w_{i,j}=1$ for all edges $e_{i,j}$, then the graph will become an unweighted graph. Let us now consider an unweighted graph $G'$ which has $n$ vertices $u_{i},~i=1,...,n$. If $A(G')$ denote the adjacency matrix of the graph $G'$ then adjacency matrix can be defined as $n \times n$ matrix $[a_{ij}]$ where $a_{ij}$ is given by
\begin{eqnarray}
a_{ij}&=&1,~~\textrm{if}~~ i \sim j
\nonumber\\&=& 0,~~\textrm{if}~~ i \nsim j
\label{intro1}
\end{eqnarray}
The Laplacian matrix $L(G')$ of the graph $G'$ may be defined as $L(G')=\Delta(G')-A(G')$, where $\Delta(G')$ denote the diagonal matrix whose diagonal entries represent the degree $d_{i}$ of each vertex $u_{i}$ of the graph $G'$. Therefore, the Laplacian matrix $L(G')=[l'_{ij}]$ of an unweighted graph $G'$ is given by
\begin{eqnarray}
l'_{ij}&=&d_{i},~~\textrm{if}~~ i = j
\nonumber\\&=& -1,~~\textrm{if}~~ i \sim j
\nonumber\\&=& 0,~~\textrm{if}~~ i \nsim j, i\neq j
\label{intro2}
\end{eqnarray}
Analogously, The Laplacian matrix $L(G)$ of the weighted graph $G$ is defined as the $n \times n$ matrix $L(G)=[l_{ij}]$, where
\begin{eqnarray}
l_{ij}&=&w_{i},~~\textrm{if}~~ i = j
\nonumber\\&=& -w_{ij},~~\textrm{if}~~ i \sim j
\nonumber\\&=& 0,~~\textrm{if}~~ i \nsim j, i\neq j
\label{intro3}
\end{eqnarray}
The Laplacian matrix $L(G)$ is a real symmetric matrix. Using the fact that $L(G)$ is a real symmetric matrix and Gershgorin's theorem, it can be shown that the eigenvalues of $L(G)$ are non-negative real numbers \cite{rojo}. Thus $L(G)$ represent a positive semi-definite matrix.\\
There exist a vast literature \cite{anderson,grone,li,merris,pan,rojo1} on Laplacian eigenvalues and their relation to various properties of the simple unweighted graph  but there exist few literatures \cite{das1,das2,poignard,chung} that studied the properties of the Laplacian matrix of the simple weighted graph. In this work, we will design the simple weighted graph corresponding to the Laplacian generated as a output of the constructed linear positive unital map. \\
\section{Tensor Formalism}
The state space of a composite quantum system can be represented as the tensor product of the state spaces of component quantum systems. For instance, if a composite quantum system contain two components and if the 1st and 2nd component are described by the Hilbert space $H_{A}$ and $H_{B}$ respectively then the composite quantum system is described by $H_{A}\otimes H_{B}$. Generalising to n components $|\psi_{1}\rangle$, $|\psi_{2}\rangle$, $|\psi_{3}\rangle$,....,$|\psi_{n}\rangle$, the state $|\psi\rangle$ of the total system is given by    
\begin{eqnarray}
	|\psi\rangle = \bigotimes_{i=1}^{n}|\psi_{i}\rangle = |\psi_{1}\rangle \otimes |\psi_{2}\rangle \otimes.....\otimes |\psi_{n}\rangle 
	\label{tf1}
\end{eqnarray}
For a bipartite system, let us assume $i_{A}$, $i=0,1,2,.....,d_{1}-1$ be a basis for the $d_{1}-dimensional$ Hilbert space $H_{A}$ and $j_{B}$, $j=0,1,2,.....,d_{2}-1$ be a basis for the $d_{2}-dimensional$ Hilbert space $H_{B}$. Then the $d_{1}d_{2}$ states $|i_{A}\rangle\otimes |j_{B}\rangle$ form a basis for the composite space $H_{AB}=H_{A}\otimes H_{B}$. Therefore, the dimension of the Hilbert space $H_{AB}$ is $d_{1}d_{2}$. For a two-qubit system, the four basis states are given by
\begin{eqnarray}
&& B_{1}\equiv |00\rangle\equiv |0\rangle \otimes |0\rangle= \begin{pmatrix}
 1 \\
 0
\end{pmatrix} \otimes \begin{pmatrix}
1 \\
0
\end{pmatrix} = \begin{pmatrix}
1 \\
0 \\
0 \\
0
\end{pmatrix}\nonumber\\&&
B_{2}\equiv |01\rangle\equiv |0\rangle \otimes |1\rangle= \begin{pmatrix}
	1 \\
	0
\end{pmatrix} \otimes \begin{pmatrix}
	0 \\
	1
\end{pmatrix} = \begin{pmatrix}
	0 \\
	1 \\
	0 \\
	0
\end{pmatrix}\nonumber\\&&
B_{3}\equiv |10\rangle\equiv |1\rangle \otimes |0\rangle= \begin{pmatrix}
	0 \\
	1
\end{pmatrix} \otimes \begin{pmatrix}
	1 \\
	0
\end{pmatrix} = \begin{pmatrix}
	0 \\
	0 \\
	1 \\
	0
\end{pmatrix}\nonumber\\&&
B_{4}\equiv |11\rangle\equiv |1\rangle \otimes |1\rangle= \begin{pmatrix}
	0 \\
	1
\end{pmatrix} \otimes \begin{pmatrix}
	0 \\
	1
\end{pmatrix} = \begin{pmatrix}
	0 \\
	0 \\
	0 \\
	1
\end{pmatrix}
\label{tf2}
\end{eqnarray}  
An arbitrary dimensional bipartite state $|\psi\rangle_{AB}\in H_{AB}$ can be expressed as
\begin{eqnarray}
|\psi\rangle_{AB}= \sum_{i,j}\alpha_{i,j}|i_{A}\rangle\otimes |j_{B}\rangle,~~\sum_{i,j}|\alpha_{i,j}|^2=1
\label{tf2}
\end{eqnarray}
The concept of bases for tensor product spaces can be extended from the arbitrary dimensional bipartite system to the multipartite system.\\
The tensor product structure is linear and it also satisfies the associative and distributive properties. But the commutative property is not satisfied by the tensor product.
\section{Preliminary Results}
\noindent In this section, we recapitulate the important results obtained in the previous works and then we will use these results in the forthcoming sections.\\
\textbf{Result-1 \cite{lasserre,kumari}:} If $X$ and $Y$ denote any two $n \times n$ Hermitian matrices then we have
\begin{eqnarray}
\lambda_{min}[X]tr[Y] \leq tr[XY] \leq \lambda_{max}[X]tr[Y]
\label{result5}
\end{eqnarray}
where $\lambda_{min}[X]$ and $\lambda_{max}[X]$ denote the minimum and maximum eigenvalue of $X$. $tr[.]$ denote the trace of the matrix $[.]$.\\
\textbf{Result-2:} Let $X,Y \in M_{n}$ be Hermitian matrices and let the $i^{th}$ eigenvalues of $X$, $Y$ and $X+Y$ be denoted by $\lambda_{i}[X]$, $\lambda_{i}[Y]$ and $\lambda_{i}[X+Y]$. For $k=1,2,3,....,n$, the Weyl's inequality \cite{horn} is given by
\begin{eqnarray}
\lambda_{k}[X]+\lambda_{min}[Y]\leq \lambda_{k}[X+Y] \leq \lambda_{k}[X]+\lambda_{max}[Y]
\label{result2}
\end{eqnarray}
where $M_{n}$ denote the set of $n \times n$ Hermitian matrices and $\lambda_{min}(Y)$, $\lambda_{max}(Y)$ denotes the minimum and maximum eigenvalues of Y, respectively.\\
\textbf{Result-3:} Let $\Phi: M_{n}\rightarrow M_{k}$ be a positive unital linear map. $M_{n}$ and $M_{k}$ denote the set of all matrices of order $n$ and $k$ respectively. For every $n \times n$ Hermitian matrix $A$, the inequality given below follows:
\begin{eqnarray}
[\Phi(A)]^{2}\leq \Phi(A^{2})
\label{result3}
\end{eqnarray}
The inequality (\ref{result3}) is known as Kadison's inequality \cite{kadison}.\\
\textbf{Result-4 \cite{wolkowicz}:} Let $C$ be an $n \times n$ complex matrix with real eigenvalues $\lambda[C]$. Then
\begin{eqnarray}
m-s\sqrt{n-1}\leq \lambda_{min}[C] \leq m-\frac{s}{\sqrt{n-1}}
\label{result4}
\end{eqnarray}
where $m=\frac{tr[C]}{n}$ and $s^{2}=\frac{tr[C^{2}]}{n}-m^{2}$.\\
\textbf{Result-5 \cite{rojo}:} If a simple connected weighted graph of order $n$ represented by $G$ then
\begin{eqnarray}
\lambda_{max}[L] \leq \frac{1}{2}max_{i\sim j}W[i,j]
\label{result5}
\end{eqnarray}
where $L$ denote the Laplacian corresponding to the graph $G$ and $W[i,j]$ is given by \cite{rojo}
\begin{eqnarray}
W[i,j]&=&w_{i}+w_{j}+\sum_{k\sim i,k \nsim j}w_{ik}+\sum_{k\sim j,k \nsim i}w_{jk}\nonumber\\&+&\sum_{k\sim i,k \sim j}|w_{ik}-w_{jk}|
\label{wij}
\end{eqnarray}

\section{Construction of a unital map}
\noindent In this section, we construct a map that may provide Laplacian at the output and also study the properties of the constructed map. Then we provide the physical interpretation of the total degree of the graph, which is constructed from the generated Laplacian from the map.\\
Let $M_{n}(R^{+})$ denote the set of all $n\times n$ matrices over the set $R^{+}$ of all positive real numbers.
The map $\phi:M_{n}(R^+)\rightarrow M_{n}(R^{+})$ may be defined as
\begin{eqnarray}
\phi (A) = L_{A}+ A
\label{map}
\end{eqnarray}
where $A\in M_{n}(R^+)$ and $L_{A}$ denote a matrix of order $n$, which is constructed using the entries of the matrix $A$.\\
Let us consider the $n \times n$ real matrix $A$, which is given by
\begin{eqnarray}
A=
\begin{pmatrix}
  a_{1,1} & a_{1,2} & ...... & a_{1,n-1} & a_{1,n} \\
  a_{2,1} & a_{2,2} & ...... & a_{2,n-1} & a_{2,n} \\
  . & . & ...... & . & . \\
  a_{n,1} & a_{n,2} & ...... & a_{n,n-1} & a_{n,n}
\end{pmatrix}, \sum_{i=1}^{n}a_{i,i}=1.
\label{matrixa}
\end{eqnarray}
The matrix $L_{A}$ can be constructed as
\begin{eqnarray}
L_{A}=
\begin{pmatrix}
  d_{1} & l_{1,2} & ...... & l_{1,n-1} & l_{1,n} \\
  l_{1,2} & d_{2} & ...... & l_{2,n-1} & l_{2,n} \\
  . & . & ...... & . & . \\
  l_{1,n-1} & l_{2,n-1} & ...... & d_{n-1} & l_{n-1,n}\\
  l_{1,n} & l_{2,n} & ...... & l_{n-1,n} & d_{n}
\end{pmatrix}
\label{matrixlapa}
\end{eqnarray}
where
$l_{i,j}=-\frac{1}{2}(a_{i,j}+a_{j,i}),~i,j=1...n-1,~i\neq j.$, $d_{i}=\sum_{j\neq i,j=1}^{n}|l_{i,j}|$.\\
In particular, we can consider the domain $D\subset M_{n}(R^+)$ as the set of all $d_{1} \otimes d_{2}$ dimensional bipartite quantum states described by the density matrices of order $d_{1}d_{2}\times d_{1}d_{2}$. For the domain $D$, the map $\phi$ can be re-expressed as
\begin{eqnarray}
\phi (\rho) = L_{\rho}+ \rho
\label{map1}
\end{eqnarray}
where the input state described by the density matrix $\rho \in D$ is given by $(\ref{matrixa})$ with $\rho=[\rho_{i,j}],i,j=1,2,....,n$ and at the output $L_{\rho}$ is given by $(\ref{matrixlapa})$ with $l_{i,j}=-\rho_{i,j},~~i\neq j; i,j=1,2...n$,~~$d_{i}=\sum_{j\neq i,j=1}^{n}|l_{i,j}|$.\\
It can be easily seen that the matrix $L_{\rho}$ satisfies the following:\\
(i) $L_{\rho}$ is symmetric and positive semi-definite.\\
(i) The smallest eigenvalue of $L_{\rho}$ is zero and (1,1,1,...,1) represent an eigenvector corresponding to the smallest eigenvalue.\\
Thus, the matrix $L_{\rho}$ act as a Laplacian corresponding to the matrix $\rho$.
\subsection{Properties of the map $\phi$}
\noindent We are now in a position to discuss the properties of the map $\phi$ defined in (\ref{map}) .\\
\textbf{P-1:} $\phi$ is a linear map.\\
\textbf{Proof:} Let $A_{1}$ and $A_{2}$ be any two matrices of same order from $M_{n}(R^+)$ and let $ \alpha_{1}, \alpha_{2}$ be any two real scalars. Let $a_{i,j}$ and $b_{i,j}$ be the $(i,j)^{th}$ entries of $A_1$ and $A_2$ respectively. Then, it is clear that
$\alpha_{1} a_{i,j}+\alpha_{2} b_{i,j}$ is the $(i,j)^{th}$ entry of $L_{\alpha_1 A_1+\alpha_2 A_2}$\\
According to the definition of the Laplacian, we have
\begin{eqnarray}
L_{ \alpha_{1}A_1+\alpha_{2}A_2}=\alpha_{1}L_{A_1}+\alpha_2 L_{A_2}
\label{s1}
\end{eqnarray}
Therefore, using (\ref{s1}), $\phi(\alpha_{1}A_1+\alpha_{2}A_{2})$ can be written as
\begin{eqnarray}
\phi(\alpha_{1}A_1+\alpha_{2}A_{2})&=& L_{\alpha_{1}A_1+\alpha_{2}A_{2}} + \alpha_{1}A_{1}+\alpha_{2}A_{2}\nonumber\\&=&
\alpha_{1}L_{A_1}+\alpha_2 L_{A_2}+ \alpha_{1}A_{1}+\alpha_{2}A_{2}
\nonumber\\&=& \alpha_{1} (L_{A_{1}}+ A_{1}) + \alpha_{2} (L_{A_{2}} + A_{2})
\nonumber\\&=& \alpha_{1}\phi(A_{1})+\alpha_{2}\phi(A_{2})
\label{linear2}
\end{eqnarray}
Hence, the map $\phi$ is linear.\\\\
\textbf{P-2:} The map $\phi$ is unital i.e. $\phi(I)=I$.\\
\textbf{Proof:} It follows from the definition of the map $\phi$.\\\\
\textbf{P-3:} If the input matrix $A$ is Hermitian and positive semi-definite then  $\phi(A)$ is also positive-semidefinite.\\
\textbf{Proof:} Let $\lambda_{min}(.)$ be minimum eigenvalue of $(.)$. Since $A$ and $L_{A}$ are Hermitian matrices so from Weyl's inequality, we have
\begin{eqnarray}
\lambda_{min}(A)+ \lambda_{min}(L_{A}) \leq \lambda_{min}(A+L_{A})=\lambda_{min}(\phi(A))
\label{weyl1}
\end{eqnarray}
Since $L_{A}$ denote the Laplacian corresponding to the density matrix $A$ so $\lambda_{min}(L_{A})=0$. Therefore, (\ref{weyl1}) reduces to
\begin{eqnarray}
\lambda_{min}(A+L_{A})=\lambda_{min}(\phi(A)) \geq \lambda_{min}(A) \geq 0
\label{weyl2}
\end{eqnarray}
The last inequality follows from the positive semi-definiteness of $A$. Hence proved.
\subsection{Physical interpretation of the total degree of the graph corresponding to the Laplacian $L_{\rho}$}
To start with, let us recall the quantum state in $d_{1} \otimes d_{2}$ dimensional space described by the density matrix $\rho$. The Laplacian corresponding to $\rho$ is denoted by $L_{\rho}$. Let $G$ be the simple weighted graph for the Laplacian $L_{\rho}$ and let $d_{G}$ be total degree of the graph $G$. Then $d_{G}$ can be expressed as
\begin{eqnarray}
d_{G}=Tr[L_{\rho}]&=& d_{1}+d_{2}+.....+d_{n}\nonumber\\&=&
\sum_{i,j,i\neq j}|\rho_{i,j}|\nonumber\\&=&
C_{l_{1}}(\rho)
\label{degree}
\end{eqnarray}
where $C_{l_{1}}(\rho)$ denote the ${l_1}$-norm of quantum coherence, which is defined as the summation of modulus of the off-diagonal terms of given quantum state $\rho$. Thus, the total degree of the graph $G$ corresponding to the Laplacian of the density matrix $\rho$ can be interpreted as the $l_{1}-$ norm of the coherence of the state $\rho$.
\section{Necessary condition for the determination of the purity of a quantum state}
A quantum system can exist in two forms: a pure state or a mixed state. A pure state is a projector while the mixed state can be expressed as a convex combination of pure states. It is not always possible to prepare a pure state in the laboratory due to noisy environment. Thus, it is an important issue for the experimentalist to ascertain whether the state prepared in the laboratory is the pure state or the mixed state. To probe this, some method is needed by which pure state and mixed state can be identified. The oldest and easiest method that can be adopted to discriminate pure and mixed state is the following:\\
(i) $Tr(\rho^{2})=1$, if the state is pure.\\
(ii) $Tr(\rho^{2})<1$, if the state $\rho$ is a mixed state.\\
We need two copies of the state to implement this method. Linear entropy is another possible way to distinguish pure and mixed state. It is a quantity that can quantify the amount of mixedness in the quantum state. The linear entropy $S_{L}$ for the $d \times d$ density matrix $\rho$ can be defined as
\begin{eqnarray}
S_{L}=\frac{d^{2}}{d^{2}-1}(1-Tr(\rho^{2}))
\label{linent}
\end{eqnarray}
In case of pure state, $S_{L}=0$ while $0 < S_{L} \leq 1$ holds for mixed state.\\
The linear entropy involves non-linear functional of the quantum state and thus the value of the linear entropy depends on the $d^{2}-1$ parameter of the quantum state. All the unknown parameters of the quantum state can be determined by tomography. The method of tomography needs lot of measurement to get the information about the state parameter and the number of measurement increases as the dimension of the system increases. Additionally, tomography is very expensive in terms of resources also. Thus, in order to bypass the procedure of tomography, Ekert et.al. \cite{ekert1} have devised the quantum network, which is controlled by input data. This method require only to estimate $d-1$ parameters to extract the information about $d^{2}-1$ parameters of $d \times d$ density matrix $\rho$. Generalised uncertainty relation can be used to discriminate pure and mixed bipartite qutrit system \cite{mal}.\\
We are now in a position to discuss the detection pure and mixed quantum system using graph theoretical approach. Our criterion depends on the linear function of $\rho$. Thus, our method requires single copy of the quantum state to test whether the state is pure or mixed. \\
\textbf{Theorem-1:} If the density operator $\rho$ represent a pure quantum state then
\begin{eqnarray}
Det(L_{\rho}+\rho-I) \leq 0
\label{th1}
\end{eqnarray}
where $Det(.)$ denote the determinant.\\ 
\textbf{Proof:} Since the density operator $\rho$ is Hermitian so using Result-3, we can re-express Kadison's inequality in terms of $\rho$ as
\begin{eqnarray}
[\Phi(\rho)]^{2}\leq \Phi(\rho^{2})
\label{result3a}
\end{eqnarray}
If the quantum state $\rho$ is pure then we have
\begin{eqnarray}
\rho^{2}=\rho
\label{th1a}
\end{eqnarray}
Combining (\ref{result3a}) and (\ref{th1a}), we get
\begin{eqnarray}
&& [\Phi(\rho)]^{2}\leq \Phi(\rho)\nonumber\\&\Rightarrow&
(L_{\rho}+\rho)^{2} \leq L_{\rho}+\rho\nonumber\\&\Rightarrow&
(L_{\rho}+\rho)(L_{\rho}+\rho-I)\leq 0 \nonumber\\&\Rightarrow&
\Phi(\rho)(L_{\rho}+\rho-I)\leq 0
\label{result3b}
\end{eqnarray}
Taking determinant both sides, we get
\begin{eqnarray}
&&Det[\Phi(\rho)(L_{\rho}+\rho-I)]\leq 0 \nonumber\\&\Rightarrow&
Det[\Phi(\rho)].Det[L_{\rho}+\rho-I]\leq 0
\label{corollary1a}
\end{eqnarray}
Since $\Phi(\rho)$ is positive so we have
\begin{eqnarray}
Det(L_{\rho}+\rho-I)\leq 0
\label{result3b}
\end{eqnarray}
Hence proved.\\
\textbf{Corollary-1:} If the state $\rho$ represent a pure state then the operator $L_{\rho}+\rho-I$ has odd number of negative eigenvalues.\\
\textbf{Illustration-1:} To illustrate Theorem-1, let us consider a two-qubit pure state $|\psi\rangle$ of the form
\begin{eqnarray}
|\psi\rangle = \frac{1}{2}|00\rangle + \frac{1}{2}|01\rangle + \frac{1}{4}|10\rangle + \sqrt{\frac{7}{16}}|11\rangle
\label{illus11}
\end{eqnarray}
The density operator $\rho_{|\psi\rangle}$ of the state $|\psi\rangle$ is given by
\begin{eqnarray}
\rho_{|\psi\rangle}=
\begin{pmatrix}
  \frac{1}{4} & \frac{1}{4} & \frac{1}{8} & \frac{\sqrt{7}}{8} \\
  \frac{1}{4} & \frac{1}{4} & \frac{1}{8} & \frac{\sqrt{7}}{8} \\
  \frac{1}{8}& \frac{1}{8} & \frac{1}{16} & \frac{\sqrt{7}}{16}\\
  \frac{\sqrt{7}}{8} & \frac{\sqrt{7}}{8} & \frac{\sqrt{7}}{16} & \frac{7}{16}
\end{pmatrix}
\label{matrixillusl1}
\end{eqnarray}
The Laplacian associated with the density matrix $\rho_{|\psi\rangle}$ is given by
\begin{eqnarray}
L_{\rho_{|\psi\rangle}}=
\begin{pmatrix}
  \frac{3}{8}+\frac{\sqrt{7}}{8} & -\frac{1}{4} & -\frac{1}{8} & -\frac{\sqrt{7}}{8} \\
  -\frac{1}{4} & \frac{3}{8}+\frac{\sqrt{7}}{8} & -\frac{1}{8} & -\frac{\sqrt{7}}{8} \\
  -\frac{1}{8}& -\frac{1}{8} & \frac{1}{4}+\frac{\sqrt{7}}{16} & -\frac{\sqrt{7}}{16}\\
 -\frac{\sqrt{7}}{8} & -\frac{\sqrt{7}}{8} & -\frac{\sqrt{7}}{16} & \frac{5\sqrt{7}}{16}
\end{pmatrix}
\label{laplacianillusl1}
\end{eqnarray}
If $I_{4}$ denote the $4 \times 4$ identity matrix then the value of the determinant of the operator $L_{\rho_{|\psi\rangle}}+\rho_{|\psi\rangle}-I_{4}$ is given by
\begin{eqnarray}
Det[L_{\rho_{|\psi\rangle}}+\rho_{|\psi\rangle}-I_{4}]= -0.00027059 < 0
\label{illus12}
\end{eqnarray}
Thus, Theorem-1 is verified. Also, one can easily verify that there are three negative eigenvalues and one positive eigenvalue of the operator $L_{\rho_{|\psi\rangle}}+\rho_{|\psi\rangle}-I_{4}$. This verify the corollary-1.\\
\textbf{Theorem-2:} If $Det(L_{\rho}+\rho-I)> 0$ then the state described by the density operator $\rho$ is a mixed state.\\
\textbf{Corollary-2:} If the operator $L_{\rho}+\rho-I$ has either all eigenvalues positive or even number of negative eigenvalues then the density operator $\rho$ represent a mixed state.\\
\textbf{Illustration-2:} Let us take a $2 \otimes 4$ dimensional bipartite quantum state described by the density operator $\rho_{1}$, which is given by
\begin{eqnarray}
\rho_{1}=
\begin{pmatrix}
  \frac{1}{8} & 0 & 0 & 0 & \frac{1}{81} & 0 & 0 & \frac{1}{81} \\
  0 & \frac{1}{8} & 0 & 0 & 0 & 0 & \frac{1}{81} & 0 \\
  0 & 0 & \frac{1}{8} & 0 & 0 & \frac{1}{8} & 0 & 0 \\
 0 & 0 & 0 & \frac{1}{8} & \frac{1}{81} & 0 & 0 & \frac{1}{81} \\
 \frac{1}{81} & 0 & 0 & \frac{1}{81} & \frac{1}{8} & 0 & 0 & 0 \\
 0 & 0 & \frac{1}{8} & 0 & 0 & \frac{1}{8} & 0 & 0\\
 0 & \frac{1}{81} & 0 & 0 & 0 & 0 & \frac{1}{8} & 0\\
 \frac{1}{81} & 0 & 0 & \frac{1}{81} & 0 & 0 & 0 & \frac{1}{8}\\
\end{pmatrix}
\label{matrixillus21}
\end{eqnarray}
The Laplacian associated with the density matrix $\rho_{1}$ is given by
\begin{eqnarray}
L_{\rho_{1}}=
\begin{pmatrix}
  \frac{2}{81} & 0 & 0 & 0 & -\frac{1}{81} & 0 & 0 & -\frac{1}{81} \\
  0 & \frac{1}{81} & 0 & 0 & 0 & 0 & -\frac{1}{81} & 0 \\
  0 & 0 & \frac{1}{8} & 0 & 0 & -\frac{1}{8} & 0 & 0 \\
 0 & 0 & 0 & \frac{2}{81} & -\frac{1}{81} & 0 & 0 & -\frac{1}{81} \\
 -\frac{1}{81} & 0 & 0 & -\frac{1}{81} & \frac{2}{8} & 0 & 0 & 0 \\
 0 & 0 & -\frac{1}{8} & 0 & 0 & \frac{1}{8} & 0 & 0\\
 0 & -\frac{1}{81} & 0 & 0 & 0 & 0 & \frac{1}{81} & 0\\
 -\frac{1}{81} & 0 & 0 & -\frac{1}{81} & 0 & 0 & 0 & \frac{2}{81}\\
\end{pmatrix}
\label{matrixillus22}
\end{eqnarray}
If $I_{8}$ denote the $8 \times 8$ identity matrix then the value of the determinant of the operator $L_{\rho_{1}}+\rho_{1}-I_{8}$ is given by
\begin{eqnarray}
Det[L_{\rho_{1}}+\rho_{1}-I_{8}]\backsimeq 0.2188278 > 0
\label{illus23}
\end{eqnarray}
From (\ref{illus23}), we can conclude that the state described by the density matrix $\rho_{1}$ is a mixed state. Moreover, we find that all eigenvalues of the operator $L_{\rho_{1}}+\rho_{1}-I_{8}$ are negative. Thus, the number of negative eigenvalues are even and hence this verify the corollary-2.\\


\section{PPT criterion in terms of laplacian for $d_{1} \otimes d_{2}$ dimensional system}
\noindent A bipartite entangled state in $d_{1} \otimes d_{2}$ dimensional system can be divided into two categories: (i) Negative partial transpose entangled states (NPTES) and (ii) Positive partial transpose entangled states (PPTES) or bound entangled states. The first criterion for the entanglement detection problem was given by Peres and Horodecki \cite{peres,horodecki2} and it may be called as PH criterion. The PH criteria states that a quantum state is separable if and only if the eigenspectrum of partial transposed state contain positive eigenvalues. The criterion is necessary and sufficient for $2\otimes 2$ and $2 \otimes 3$ system but in higher dimensional systems, there exist entangled states which satisfy the PH criterion. This means that the eigenspectrum of partial transposition of the density matrix that represent the entangled state contain positive eigenvalues. The states which possesses this type of properties are known as positive partial transpose entangled states (PPTES) or bound entangled states (BES). Since PPTES are not detected by partial transposition method so other criterion such as the computable cross norm and re-alignment Criterion \cite{rudolph,chen}, range criterion \cite{phorodecki}, majorization criterion \cite{nk} developed in detecting the bound entangled states. D. P. DiVincenzo et.al. \cite{dpd} provided an example of a class of bipartite bound entangled state in $d\otimes d$ system which is also negative partial transpose entangled state.\\
In this section, we derive few criterion based on (i) the spectrum of the density matrix of the state under investigation and (ii) the spectrum of the Laplacian corresponding to the density matrix. The criterion may serve to identify PPT states and also take part in detecting the negative partial transpose entangled states (NPTES). We then illustrate our criterion by taking few examples of quantum states in higher dimensional system.

\subsection{Few PPT Criterion}
\noindent Firstly, we will derive the separability criterion for a bipartite quantum state which is proved to be necessary and sufficient in $2 \otimes 2$ dimensional system. Then we will show that the condition is only sufficient for the $d_{1} \otimes d_{2}$ dimensional bipartite PPT state where either $d_{1}, d_{2} \geq 3$ or $d_{1}=2, d_{2} > 3$. Secondly, we deduce PPT criterion for $d_{1} \otimes d_{2}$ dimensional bipartite state in terms of the minimum eigenvalue of the probe state and its corresponding Laplacian.\\
\textbf{Theorem-3:} The state described by the density operator $\rho_{AB}^{2 \otimes 2}$ in $2 \otimes 2$ dimensional system is a separable state if and only if
\begin{eqnarray}
\lambda_{min}[L_{\rho_{AB}^{2 \otimes 2}}+(\rho_{AB}^{2 \otimes 2})^{T_{B}}] \geq 0
\label{th3}
\end{eqnarray}
where $T_{B}$ denote the partial transposition with respect to the system $B$ and $L_{\rho_{AB}^{2 \otimes 2}}$ represent the laplacian corresponding to the density operator $\rho_{AB}^{2 \otimes 2}$.\\
\textbf{Proof:} For $2 \otimes 2$ dimensional system, the state $\rho_{AB}^{2 \otimes 2}$ is separable if and only if $(\rho_{AB}^{2 \otimes 2})^{T_{B}} \geq 0$. This implies
\begin{eqnarray}
\lambda_{min}[(\rho_{AB}^{2 \otimes 2})^{T_{B}}] \geq 0
\label{prth31}
\end{eqnarray}
Using the Result-2 given in (\ref{result2}) on two Hermitian operators $\rho_{AB}$ and $L_{\rho_{AB}}$, we get
\begin{eqnarray}
\lambda_{min}[L_{\rho_{AB}^{2 \otimes 2}}+(\rho_{AB}^{2 \otimes 2})^{T_{B}}] \geq \lambda_{min}[L_{\rho_{AB}^{2 \otimes 2}}]+\lambda_{min}[(\rho_{AB}^{2 \otimes 2}]^{T_{B}}]
\label{prth32}
\end{eqnarray}
Since $L_{\rho_{AB}^{2 \otimes 2}}$ is the Laplacian so $\lambda_{min}[L_{\rho_{AB}^{2 \otimes 2}}]=0$. Therefore, the inequality (\ref{prth32}) reduces to
\begin{eqnarray}
\lambda_{min}[L_{\rho_{AB}^{2 \otimes 2}}+(\rho_{AB}^{2 \otimes 2})^{T_{B}}] \geq \lambda_{min}[(\rho_{AB}^{2 \otimes 2})^{T_{B}}]
\label{prth33}
\end{eqnarray}
Using (\ref{prth31}) and (\ref{prth33}), we get the required result.\\
One may note that the necessary and sufficient condition given in Theorem-3 also holds for $2 \otimes 3$ dimensional bipartite system but the condition is only sufficient in the higher dimensional system because of the existence of positive partial transpose entangled states (PPTES). Thus, theorem-3 may be modified for the higher dimensional system in the following way:\\
\textbf{Theorem-4:} If any bipartite state in $d_{1} \otimes d_{2}$ (either $d_{1},d_{2} \geq 3$ or $d_{1}=2,d_{2}> 3$), dimensional system described by the density operator $\rho_{AB}^{d_{1} \otimes d_{2}}$ represent a positive partial transpose state then it satisfies the inequality
\begin{eqnarray}
\lambda_{min}(L_{\rho_{AB}^{d_{1} \otimes d_{2}}}+(\rho_{AB}^{d_{1} \otimes d_{2}})^{T_{B}}) \geq 0
\label{th4}
\end{eqnarray}
We note here that the condition (\ref{th4}) is only sufficient for $d_{1} \otimes d_{2}$ dimensional system while the condition (\ref{th3}) is necessary and sufficient for $2 \otimes 2$ and $2 \otimes 3$ dimensional system.\\
\textbf{Corollary-4:} If any arbitrary $d_{1} \otimes d_{2}$ dimensional bipartite state $\rho_{AB}^{d_{1} \otimes d_{2}}$ satisfies the inequality
\begin{eqnarray}
\lambda_{min}(L_{\rho_{AB}^{d_{1} \otimes d_{2}}}+(\rho_{AB}^{d_{1} \otimes d_{2}})^{T_{B}}) < 0
\label{cor4}
\end{eqnarray}
then the state $\rho_{AB}^{d_{1} \otimes d_{2}}$ is negative partially transposed entangled state (NPTES).\\
Next our task is to derive few other criterion based on the maximum and minimum eigenvalues of the Laplacian of the given state that may identify whether the given state is a PPT state or not? These conditions are necessary conditions and hence they are of particular importance.\\
\textbf{Theorem-5:} If any arbitrary full rank $d_{1} \otimes d_{2}$ dimensional state described by the density operator $\rho$ satisfies the inequality
\begin{eqnarray}
\lambda_{min}(\rho) \geq \lambda_{max}(L_{\rho}^{T_{B}})-\lambda_{min}(L_{\rho}^{T_{B}})
\label{th5}
\end{eqnarray}
then the state $\rho$ is either separable state or PPTES.\\
\textbf{Proof:} To start with, let us consider the functional $Tr[(L_{\rho}+\rho)\rho^{T_{B}}]$.
\begin{eqnarray}
Tr[(L_{\rho}+\rho)\rho^{T_{B}}]&=&Tr[L_{\rho}\rho^{T_{B}}]+Tr[\rho\rho^{T_{B}}]\nonumber\\&=&Tr[L_{\rho}^{T_{B}}\rho]+Tr[\rho\rho^{T_{B}}]
\nonumber\\&\geq& \lambda_{min}[L^{T_{B}}]+\lambda_{min}[\rho]
\label{prth51}
\end{eqnarray}
We may now proceed further with the inequality (\ref{prth51}) that may be re-expressed as
\begin{eqnarray}
\lambda_{min}[L^{T_{B}}]+\lambda_{min}[\rho]&\leq& Tr[(L_{\rho}+\rho)\rho^{T_{B}}]\nonumber\\&=& Tr[(L_{\rho}+\rho)^{T_{B}}\rho]\nonumber\\&=&Tr[(L_{\rho}^{T_{B}}+\rho^{T_{B}})\rho]\nonumber\\&\leq& \lambda_{max}[L^{T_{B}}+\rho^{T_{B}}]\nonumber\\&\leq& \lambda_{max}[L^{T_{B}}]+\lambda_{min}(\rho^{T_{B}})
\label{prth52}
\end{eqnarray}
Rearranging the terms of the inequality (\ref{prth52}), we get
\begin{eqnarray}
\lambda_{min}[\rho^{T_{B}}]&\geq& \lambda_{min}[\rho]-[\lambda_{max}[L^{T_{B}}]-\lambda_{min}[L^{T_{B}}]]
\label{prth53}
\end{eqnarray}
Therefore, if a quantum state $\rho$ satisfies the inequality $\lambda_{min}[\rho]\geq \lambda_{max}[L^{T_{B}}]-\lambda_{min}[L^{T_{B}}]$, then $\lambda_{min}[\rho^{T_{B}}] \geq 0$. Thus, the state $\rho$ represent either a separable state or bound entangled state. Hence proved.\\
\textbf{Corollary-5:} If the inequality (\ref{th5}) is violated by any quantum state $\rho$ then the state may or may not be NPTES.\\
Now it can be seen that the criterion given in (\ref{th5}) depends on the partial transposition of the Laplacian of the given state but since partial transposition is not a physical operation so it cannot be implemented in the laboratory. Therefore, the above theorem and corollary that involve partial transposition operation, cannot be used in an experiment for the detection of positive partial transpose states. Thus, we need to modify Theorem-5 in such a way so that we can avoid partial transposition operation in deducing the condition for the detection of states with positive partial transpose. In doing this, we will deduce another criterion which is free from partial transposition operation but can discriminate between positive partial transpose states and negative partial transpose states. The modified criterion can be expressed via the following theorem:\\
\textbf{Theorem-6:}  If a full rank state $\rho$ that exist in $d_{1} \otimes d_{2}$ dimensional system satisfies the inequality
\begin{eqnarray}
\lambda_{min}[\rho] \geq \lambda_{max}[L_{\rho}]
\label{th6}
\end{eqnarray}
then the state $\rho$ is either separable state or PPTES.\\
\textbf{Proof:} Let us begin with the functional $Tr[(L_{\rho}+\rho^{T_{B}})\rho]$. The lower bound of $Tr[(L_{\rho}+\rho^{T_{B}})\rho]$ can be derived as
\begin{eqnarray}
Tr[(L_{\rho}+\rho^{T_{B}})\rho]&=&Tr[L_{\rho}\rho]+Tr[\rho^{T_{B}}\rho]\nonumber\\&\geq& \lambda_{min}[L_{\rho}]+\lambda_{min}(\rho)
\label{prth61}
\end{eqnarray}
The inequality (\ref{prth61}) may be written as
\begin{eqnarray}
\lambda_{min}[L_{\rho}]+\lambda_{min}[\rho] &\leq& Tr[(L_{\rho}+\rho^{T_{B}})\rho]\nonumber\\&\leq& \lambda_{max} [L_{\rho}+\rho^{T_{B}}]
\nonumber\\&\leq& \lambda_{max}[L_{\rho}]+\lambda_{min}[\rho^{T_{B}}]
\label{prth62}
\end{eqnarray}
Considering the fact that $\lambda_{min}[L_{\rho}]=0$ and then rearranging the terms of the inequality (\ref{prth62}), we get
\begin{eqnarray}
\lambda_{min}[\rho^{T_{B}}]&\geq& \lambda_{min}[\rho]-\lambda_{max}[L_{\rho}]
\label{prth63}
\end{eqnarray}
If we now further consider that $\lambda_{min}[\rho] \geq \lambda_{max}[L_{\rho}]$ then $\lambda_{min}[\rho^{T_{B}}]\geq 0$ and thus we can conclude that the state under investigation will be either separable state or bound entangled state, that is, a positive partial transpose state.
\subsection{Examples}
\textbf{Example-1:} Let us consider a $2 \otimes 2$ dimensional bipartite system $AB$ described by the density operator $\rho_{AB}$. It is given by
\begin{eqnarray}
\rho_{AB}=
\begin{pmatrix}
  0.1 & 0 & 0 & 0 \\
  0 & 0.2 & x & 0 \\
  0 & x & 0.4 & 0 \\
  0 & 0 & 0 & 0.3
\end{pmatrix}, 0 \leq x \leq 0.283
\label{ex1a}
\end{eqnarray}
If $\rho_{AB}^{T_{B}}$ denote the partial transposition of the state $\rho_{AB}$ then the eigenvalues of $\rho_{AB}^{T_{B}}$ are given by
\begin{eqnarray}
&&\lambda_{1}[\rho_{AB}^{T_{B}}]=0.4,\lambda_{2}[\rho_{AB}^{T_{B}}]=0.2,\nonumber \\&&
\lambda_{3}[\rho_{AB}^{T_{B}}]=\frac{1}{5}+\frac{1}{10}\sqrt{1+100x^{2}},
\nonumber\\&& \lambda_{4}[\rho_{AB}^{T_{B}}]=\frac{1}{5}-\frac{1}{10}\sqrt{1+100x^{2}}
\label{ex1b}
\end{eqnarray}
The minimum eigenvalue of $\rho_{AB}^{T_{B}}$ is given by
\begin{eqnarray}
\lambda_{min}[\rho_{AB}^{T_{B}}]=\frac{1}{5}-\frac{1}{10}\sqrt{1+100x^{2}} &\geq& 0, \textrm{for}~~ 0 \leq x \leq 0.173
\nonumber\\&\leq& 0, \textrm{for}~~ 0.173 < x \leq 0.283\nonumber\\
\label{ex1c}
\end{eqnarray}
Therefore, the state $\rho_{AB}$ is separable for $0 \leq x \leq 0.173$ while it is entangled for $0.173 < x \leq 0.283$.
If $L_{\rho_{AB}}$ denote the Laplacian associated with the density matrix $\rho_{AB}$ then it is given by
\begin{eqnarray}
L_{\rho_{AB}}=
\begin{pmatrix}
  0 & 0 & 0 & 0 \\
  0 & x & -x & 0 \\
  0 & -x & x & 0 \\
  0 & 0 & 0 & 0
\end{pmatrix}, 0 \leq x \leq 0.283
\label{ex1d}
\end{eqnarray}
The eigenvalues of $L_{\rho_{AB}}$ are given by $0,0,0,2x$. The eigenvalues of the partial transposition of $L_{\rho_{AB}}$, which is denoted by $L_{\rho_{AB}}^{T_{B}}$ are given by $-x,x,x,x$.\\
Now we are in a position to make the following observation:\\
\textbf{Observation-1:} The minimum eigenvalue of $L_{\rho_{AB}}+\rho_{AB}^{T_{B}}$ is given by $\lambda_{min}[L_{\rho_{AB}}+\rho_{AB}^{T_{B}}]=\frac{1}{5}-\frac{1}{10}\sqrt{1+100x^{2}}$, which is non-negative for $0\leq x \leq 0.173$. The region $0\leq x \leq 0.173$ represent the separability region and thus Theorem-3 is satisfied for the state $\rho_{AB}$.\\
\textbf{Observation-2:} $\lambda_{min}[L_{\rho_{AB}}+\rho_{AB}^{T_{B}}]=\frac{1}{5}-\frac{1}{10}\sqrt{1+100x^{2}}<0$ for $0.173 < x \leq 0.283$. Therefore, the region $0.173 < x \leq 0.283$ represent the entanglement region and thus Corollary-4 is satisfied for the state $\rho_{AB}$.\\
\textbf{Observation-3:} The rank of $\rho_{AB}$ given in (\ref{ex1a}) is 4. Thus, the state $\rho_{AB}$ is a full rank state. It can be easily verified that Theorem-5 and Theorem-6 are satisfied for $0 \leq x \leq 0.05$.\\
\textbf{Example-2:} Let us now consider a $2 \otimes 4$ dimensional bipartite system described by the density operator $\rho_{2}$. It is given by \cite{adhikari}
\begin{eqnarray}
\rho_{2}=
\begin{pmatrix}
  \frac{1}{8} & 0 & 0 & 0 & \frac{1}{81} & 0 & 0 & \frac{1}{81} \\
  0 & \frac{1}{8} & 0 & 0 & 0 & 0 & 0 & 0 \\
  0 & 0 & \frac{1}{8} & 0 & 0 & 0 & 0 & 0 \\
  0 & 0 & 0 & \frac{1}{8} & \frac{1}{81} & 0 & 0 & \frac{1}{81} \\
  \frac{1}{8} & 0 & 0 & \frac{1}{81} & \frac{1}{8} & 0 & 0 & 0 \\
  0 & 0 & 0 & 0 & 0 & \frac{1}{8} & 0 & 0 \\
  0 & 0 & 0 & 0 & 0 & 0 & \frac{1}{8} & 0 \\
  \frac{1}{81} & 0 & 0 & \frac{1}{81} & 0 & 0 & 0 & \frac{1}{8} \\
\end{pmatrix},
\label{ex2a}
\end{eqnarray}
The eigenvalues of $\rho_{2}$ are given by
\begin{eqnarray}
&&\mu_{1}=\frac{97}{648}>\mu_{2}=\mu_{3}=\mu_{4}=\mu_{5}=\mu_{6}=\mu_{7}=\frac{1}{8}\geq \nonumber\\&&\mu_{8}=\frac{65}{648}
\label{ex2b}
\end{eqnarray}
In this example, we can find that the partial transposed state $\rho_{2}^{T_{B}}$ is identical with the state $\rho_{2}$. Thus, we have
\begin{eqnarray}
\rho_{2}=\rho_{2}^{T_{B}}
\label{ex2c}
\end{eqnarray}
It has been shown that the state $\rho_{2}$ is a separable state \cite{adhikari}.\\
If $L_{\rho_{2}}$ and $L_{\rho_{2}}^{T_{B}}$ denote the Laplacian and partial transposition of the Laplacian associated with the density matrix $\rho_{2}$ then they are given by
\begin{eqnarray}
L_{\rho_{2}}=L_{\rho_{2}}^{T_{B}}=
\begin{pmatrix}
  \frac{2}{81} & 0 & 0 & 0 & \frac{-1}{81} & 0 & 0 & \frac{-1}{81} \\
  0 & 0 & 0 & 0 & 0 & 0 & 0 & 0 \\
  0 & 0 & 0 & 0 & 0 & 0 & 0 & 0 \\
  0 & 0 & 0 & \frac{2}{81} & \frac{-1}{81} & 0 & 0 & \frac{-1}{81} \\
  \frac{-1}{81} & 0 & 0 & \frac{-1}{81} & \frac{2}{81} & 0 & 0 & 0 \\
  0 & 0 & 0 & 0 & 0 & 0 & 0 & 0 \\
  0 & 0 & 0 & 0 & 0 & 0 & 0 & 0 \\
  \frac{-1}{81} & 0 & 0 & \frac{-1}{81} & 0 & 0 & 0 & \frac{2}{81} \\
\end{pmatrix},
\label{ex2d}
\end{eqnarray}
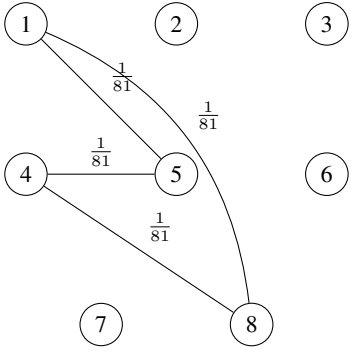
\begin{figure}[h]
	\begin{tikzpicture}
	\centering
	\node [circle, draw](1) at (0,0){1};
	\node [circle, draw](2) at (2,0){2};
	\node [circle, draw](3) at (4,0){3};
	\node [circle, draw](4) at (0,-2){4};
	\node [circle, draw](5) at (2, -2){5};
	\node [circle, draw](6) at (4,-2){6};
	\node [circle, draw](7) at (1,-4){7};
	\node [circle, draw](8) at (3,-4){8};
	
	\draw (1) edge [pos=0.5, "$\frac{1}{81}$"] (5);
	\draw (1) edge [bend left, pos=0.5,  "$\frac{1}{81}$"] (8);
	\draw (4) edge [pos=0.5, "$\frac{1}{81}$"] (5);
	\draw (4) edge [pos=0.5,  "$\frac{1}{81}$"] (8);

	\end{tikzpicture}
	\caption{Graph corresponding to $L_{\rho_{2}}$}
	\label{fig:rho2}
\end{figure}
Therefore, the eigenvalues of $L_{\rho_{2}}=L_{\rho_{2}}^{T_{B}}$ are given by $0,0,0,0,0,\frac{2}{81},\frac{2}{81},\frac{4}{81}$.\\
Now we observe the following facts:\\
\textbf{Observation-1:} It can be easily verified that the minimum eigenvalue $\lambda_{min}[L_{\rho_{2}}+\rho_{2}^{T_{B}}]$ is greater than zero. Thus according to the Theorem-4, we can infer that the state $\rho_{2}$ is a PPT state. But using Theorem-4, we are unable to tell that whether the state $\rho_{2}$ is separable state or PPTES.\\
\textbf{Observation-2:} The state $\rho_{2}$ represent a full rank state. Now, going through the eigenvalues of $\rho_{2}$ given in (\ref{ex2b}), we can determine that $\lambda_{min}[\rho_{2}]=\frac{65}{648}$. Thus, we have
\begin{eqnarray}
\lambda_{min}[\rho_{2}]=\frac{65}{648}> \lambda_{max}[L_{\rho_{2}}^{T_{B}}]-\lambda_{min}[L_{\rho_{2}}^{T_{B}}]=\frac{4}{81}
\label{ex2e}
\end{eqnarray}
Hence, Theorem-5 is satisfied for the state $\rho_{2}$.\\
\textbf{Observation-3:} Since the state $\rho_{2}$ is a full rank state so we can apply Theorem-6. We can then easily verify Theorem-6 for the state $\rho_{2}$.\\

\section{Graphical Interpretation}
\noindent Here, we will discuss the graphical interpretation of results given in section-VI A. We will show that PPT criterion of a quantum state can be interpreted through the properties of a graph associated with the quantum states.\\
\textbf{Theorem-3A:} The state described by the density operator $\rho^{2 \otimes 2}$ in $2 \otimes 2$ dimensional system is a separable state if and only if
\begin{eqnarray}
0 \leq \lambda_{min}[L_{\rho^{2 \otimes 2}}+(\rho^{2 \otimes 2})^{T_{B}}] \leq 1+d_{G}
\label{th3a}
\end{eqnarray}
where $d_{G}$ denote the total degree of the graph $G$ associated with the density operator $\rho^{2 \otimes 2}$.\\
\textbf{Proof:} The lower bound of $\lambda_{min}[L_{\rho^{2 \otimes 2}}+(\rho^{2 \otimes 2})^{T_{B}}]$ may be verified from the inequality (\ref{th3}) given in Theorem-3. Now our task is to derive the upper bound. The state $\rho^{2 \otimes 2}$ in $2 \otimes 2$ dimensional system is a separable state if and only if its partial transposed form denoted by $(\rho^{2 \otimes 2})^{T_{B}}$ is a positive semi-definite operator. Further, since $L_{\rho^{2 \otimes 2}}$ is also a positive semi-definite operator so we can write
\begin{eqnarray}
Tr[L_{\rho^{2 \otimes 2}}+(\rho^{2 \otimes 2})^{T_{B}}]\geq \lambda_{min}[L_{\rho^{2 \otimes 2}}+(\rho^{2 \otimes 2})^{T_{B}}]
\label{th3apf3a3}
\end{eqnarray}
The inequality (\ref{th3apf3a3}) can be further re-expressed as
\begin{eqnarray}
&&Tr[L_{\rho^{2 \otimes 2}}]+Tr[(\rho^{2 \otimes 2})^{T_{B}}] \geq \lambda_{min}[L_{\rho^{2 \otimes 2}}+(\rho^{2 \otimes 2})^{T_{B}}] \nonumber\\&\Rightarrow&
d_{G}+1 \geq \lambda_{min}[L_{\rho^{2 \otimes 2}}+(\rho^{2 \otimes 2})^{T_{B}}]
\label{th3apf3a4}
\end{eqnarray}
where we used the fact that $Tr[(\rho^{2 \otimes 2})^{T_{B}}]=1$ and $Tr[L_{\rho^{2 \otimes 2}}]=d_{G}$.\\
Combining Theorem-3 and the inequality (\ref{th3apf3a4}), we get the required result.\\
\textbf{Theorem-3B:} If $\rho$ denote $d_{1} \otimes d_{2}$ dimensional NPTES and if the graph $G$ corresponding to  the density matrix $\rho$ is simple connected weighted graph then
\begin{eqnarray}
\lambda_{min}[L_{\rho}+\rho^{T_{B}}] \leq \frac{1}{2} max_{i \sim j}W[i,j]
\label{th3b}
\end{eqnarray}
where $W[i,j]$ is given by Result-5.\\
\textbf{Proof:} Let us start with $\lambda_{min}[L_{\rho}+\rho^{T_{B}}]$. Using Result-2, we get
\begin{eqnarray}
\lambda_{min}[L_{\rho}+\rho^{T_{B}}] &\leq& \lambda_{max}[L_{\rho}]+\lambda_{min}[\rho^{T_{B}}]\nonumber\\&\leq& \lambda_{max}[L_{\rho}]
\nonumber\\&\leq&  \frac{1}{2} max_{i \sim j}W[i,j]
\label{thpf3b1}
\end{eqnarray}
The second inequality follows from the fact that the state $\rho$ is a NPTES and the last inequality follows from Result-5. Hence proved.\\
\textbf{Illustration-3:} Let us consider a state $\rho_{3}$, which is given by
\begin{eqnarray}
\rho_{3}=
\begin{pmatrix}
  0.4 & 0.2 & 0.1 & 0.1 \\
  0.2 & 0.3 & 0.2 & 0.1 \\
  0.1 & 0.2 & 0.2 & 0.1 \\
  0.1 & 0.1 & 0.1 & 0.1
\end{pmatrix}
\label{illus3rho}
\end{eqnarray}
The partial transposed state $\rho_{3}^{T_{B}}$ can be expressed as
\begin{eqnarray}
\rho_{3}^{T_{B}}=
\begin{pmatrix}
  0.4 & 0.2 & 0.1 & 0.2 \\
  0.2 & 0.3 & 0.1 & 0.1 \\
  0.1 & 0.1 & 0.2 & 0.1 \\
  0.2 & 0.1 & 0.1 & 0.1
\end{pmatrix}
\label{illus3rhopt}
\end{eqnarray}
The eigenvalues of $\rho_{3}^{T_{B}}$ are: $0.7,0.16,0.14,-0.01$. This shows that the state $\rho_{3}$ is a NPTES.\\
The Laplacian of the state described by the density operator $\rho_{3}$ is given by
\begin{eqnarray}
L_{\rho_{3}}=
\begin{pmatrix}
  0.4 & -0.2 & -0.1 & -0.1 \\
  -0.2 & 0.5 & -0.2 & -0.1 \\
  -0.1 & -0.2 & 0.4 & -0.1 \\
  -0.1 & -0.1 & -0.1 & 0.3
\end{pmatrix}
\label{illus3lap}
\end{eqnarray}
If $\lambda_{min}(L_{\rho_{3}}+\rho_{3}^{T_{B}})$ denote the minimum eigenvalue of $L_{\rho_{3}}+\rho_{3}^{T_{B}}$ then
\begin{eqnarray}
\lambda_{min}(L_{\rho_{3}}+\rho_{3}^{T_{B}})=0.3763
\label{illus3lambdamin}
\end{eqnarray}
\begin{figure}[h]
	\begin{tikzpicture}[node distance= 3cm ]
	\node [circle, draw](00) at (0,0){1};
	\node [circle, draw](01) at (3,0){2};
	\node [circle, draw](10)at (0,-3){3};
	\node [circle, draw](11) at (3,-3){4};
	\draw (00) --node[above =0.10 cm]{0.2}(01);
	\draw (00) --node[left =0.10 cm]{0.1}(10);
	\draw (00) edge [pos=0.25, "0.1"] (11);
	\draw (01) edge [pos=0.75, "0.2"] (10);

	\draw (01) --node[right =0.10 cm]{0.1}(11);
	\draw (10) --node[below =0.10 cm]{0.1}(11);
	\end{tikzpicture}
	\caption{Graph corresponding to $L_{\rho_{3}}$}
	\label{fig:rho3}
\end{figure}
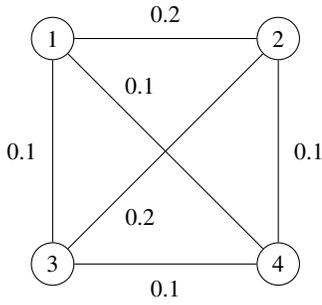

Using (\ref{wij}), we can calculate $max_{i\sim j}W[i,j]$ for the graph shown in \figref{rho3} and it is given by
\begin{eqnarray}
max_{i\sim j}W[i,j]=0.9
\label{illus3wij}
\end{eqnarray}
Therefore, the inequality (\ref{th3b}) can be easily verified using (\ref{illus3lambdamin}) and (\ref{illus3wij}).\\
\textbf{Theorem-4A:} For any $d_{1} \otimes d_{2}$ dimensional positive partial transposed bipartite state $\rho^{d_{1}\otimes d_{2}}$, the inequality
\begin{eqnarray}
1+d_{G}\geq \lambda_{min}[L_{\rho^{d_{1} \otimes d_{2}}}+(\rho^{d_{1} \otimes d_{2}})^{T_{B}}]
\label{th4a}
\end{eqnarray}
holds.\\
\textbf{Corollary-4A:} Let $d_{G}$ denote the total degree of the simple connected weighted graph $G$ associated with any arbitrary $d_{1} \otimes d_{2}$ dimensional bipartite state $\varrho^{d_{1}\otimes d_{2}}$. If the state $\varrho^{d_{1}\otimes d_{2}}$ satisfies the inequality
\begin{eqnarray}
1+d_{G} < (d_{1}d_{2}-1)(\frac{1}{2} max_{i \sim j}W[i,j] +\lambda_{max}[(\varrho^{d_{1}\otimes d_{2}})^{T_{B}}])\nonumber\\
\label{cor4a}
\end{eqnarray}
then the state $\varrho^{d_{1}\otimes d_{2}}$ is negative partially transposed entangled state (NPTES).\\
\textbf{Proof:} Let us recall corollary-4. It implies that if
\begin{eqnarray}
\lambda_{min}[L_{\varrho^{d_{1}\otimes d_{2}}}+(\varrho^{d_{1}\otimes d_{2}})^{T_{B}}]<0
\label{cor4apf4a1}
\end{eqnarray}
then the state $\varrho^{d_{1}\otimes d_{2}}$ is NPTES.\\
Using Result-4, the inequality (\ref{cor4apf4a1}) can be re-expressed as
\begin{eqnarray}
m < s\sqrt{d_{1}d_{2}-1}
\label{cor4apf4a2}
\end{eqnarray}
where $m= \frac{Tr[L_{\varrho^{d_{1}\otimes d_{2}}}+(\varrho^{d_{1}\otimes d_{2}})^{T_{B}}]}{d_{1}d_{2}}$ and $s^{2}=\frac{Tr[(L_{\varrho^{d_{1}\otimes d_{2}}}+(\varrho^{d_{1}\otimes d_{2}})^{T_{B}})^{2}]-(\frac{Tr[L_{\varrho^{d_{1}\otimes d_{2}}}+(\varrho^{d_{1}\otimes d_{2}})^{T_{B}}]}{d_{1}d_{2}})^{2}}{d_{1}d_{2}}$.\\
Simplifying the inequality (\ref{cor4apf4a2}), we get
\begin{eqnarray}
&&(1+d_{G})^{2}<(d_{1}d_{2}-1) Tr[(L_{\varrho^{d_{1}\otimes d_{2}}}+(\varrho^{d_{1}\otimes d_{2}})^{T_{B}})^{2}]\nonumber\\&\Rightarrow&
(1+d_{G})^{2}<(d_{1}d_{2}-1)\lambda_{max}[L_{\varrho^{d_{1}\otimes d_{2}}}+(\varrho^{d_{1}\otimes d_{2}})^{T_{B}}](1+d_{G})\nonumber\\&\Rightarrow&
1+d_{G}<(d_{1}d_{2}-1)\lambda_{max}[L_{\varrho^{d_{1}\otimes d_{2}}}+(\varrho^{d_{1}\otimes d_{2}})^{T_{B}}]
\label{cor4apf4a3}
\end{eqnarray}
The second line follows from Result-1. Moreover, using Result-2 on $\lambda_{max}[L_{\varrho^{d_{1}\otimes d_{2}}}+(\varrho^{d_{1}\otimes d_{2}})^{T_{B}}]$, the inequality (\ref{cor4apf4a3}) further reduces to
\begin{eqnarray}
1+d_{G} &<& (d_{1}d_{2}-1)(\lambda_{max}[L_{\varrho^{d_{1}\otimes d_{2}}}]+\lambda_{max}[(\varrho^{d_{1}\otimes d_{2}})^{T_{B}}])\nonumber\\ &\leq& (d_{1}d_{2}-1)(\frac{1}{2} max_{i \sim j}W[i,j] +\lambda_{max}[(\varrho^{d_{1}\otimes d_{2}})^{T_{B}}])\nonumber\\
\label{cor4a}
\end{eqnarray}
The last inequality follows from Result-5.\\

\textbf{Theorem-7:} If any $d_{1} \otimes d_{2}$ dimensional bipartite NPTES state described by the density operator $\rho$ then
\begin{eqnarray}
\lambda_{min}[\rho] \leq \frac{1}{2} max_{i \sim j}W[i,j]
\label{th7}
\end{eqnarray}
where $W[i,j]$ is given by Result-5 for the simple connected weighted graph $G$ associated with the density operator $\rho$.\\
\textbf{Proof:} Let us recall the inequality (\ref{prth63}), which can be re-expressed in the form as
\begin{eqnarray}
\lambda_{min}[\rho] \leq \lambda_{max}[L_{\rho}]+\lambda_{min}[\rho^{T_{B}}]
\label{prth71}
\end{eqnarray}
Since the state $\rho$ represent a NPTES so $\lambda_{min}[\rho^{T_{B}}]<0$. Thus, we have
\begin{eqnarray}
\lambda_{min}[\rho] \leq \lambda_{max}[L_{\rho}]
\label{prth72}
\end{eqnarray}
The non-trivial upper bound of $\lambda_{max}[L_{\rho}]$ is given by the Result-5. Using Result-5, the inequality (\ref{prth72}) reduces to
\begin{eqnarray}
\lambda_{min}[\rho] \leq \frac{1}{2} max_{i \sim j}W[i,j]
\label{prth73}
\end{eqnarray}
Hence proved.\\
\textbf{Corollary-6:} If any $d_{1} \otimes d_{2}$ dimensional full rank bipartite state described by the density operator $\rho$ satisfies the inequality
\begin{eqnarray}
\lambda_{min}[\rho] > \frac{1}{2} max_{i \sim j}W[i,j]
\label{cor-6}
\end{eqnarray}
then the full rank state $\rho$ must be a PPT state.\\
\textbf{Illustration-4:} The quantum state described by the density operator $\rho_{5}$ is given by
\begin{eqnarray}
\rho_{5}=
\begin{pmatrix}
  \frac{1}{4} & \frac{1}{20} & 0 & \frac{1}{20} \\
    \frac{1}{20} & \frac{1}{4} & 0 & 0 \\
  0 & 0 & \frac{1}{4} &  \frac{1}{20}  \\
  \frac{1}{20} & 0 &  \frac{1}{20}  & \frac{1}{4}
\end{pmatrix}
\label{illus5rho}
\end{eqnarray}
The eigenvalues of $\rho_{5}$ are as follows: $0.3309, 0.2809, 0.2191, 0.1691$.
The Laplacian corresponding to the density matrix is given by
\begin{eqnarray}
L_{\rho_{5}}=
\begin{pmatrix}
  \frac{1}{10} &  -\frac{1}{20}  & 0 & -\frac{1}{20} \\
  - \frac{1}{20}  &  \frac{1}{20}  & 0 & 0 \\
  0 & 0 &  \frac{1}{20}  & - \frac{1}{20}  \\
  -\frac{1}{20} & 0 & - \frac{1}{20}  & \frac{1}{10}
\end{pmatrix}
\label{illus5lap}
\end{eqnarray}
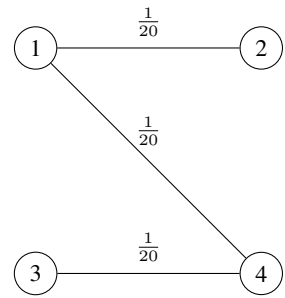
\begin{figure}[h]
	\begin{tikzpicture}[node distance= 3cm ]
	\centering
	\node [circle, draw](00) at (0,0){1};
	\node [circle, draw](01) at (3,0){2};
	\node [circle, draw](10)at (0,-3){3};
	\node [circle, draw](11) at (3,-3){4};
	\draw (00) --node[above =0.05 cm]{$\frac{1}{20}$}(01);
	\draw (00) --node[above=0.10 cm]{$\frac{1}{20}$}(11);
	\draw (10) --node[above =0.05 cm]{$\frac{1}{20}$}(11);
	\end{tikzpicture}
	\caption{Graph corresponding to $L_{\rho_{5}}$}
	\label{fig:rho5}
\end{figure}

From the graph given in Fig. 3, it is clear that there are three edges namely $e_{1,2}, e_{1,4}$ and $e_{3,4}$. Thus, we need to calculate $W[1,2], W[1,4]$ and $W[3,4]$ as per the prescription given in (\ref{wij}). Therefore, we have

\begin{eqnarray}
W[1,2]&=& w_{1}+ w_{2}+\sum_{k\sim 1,k \nsim 2}w_{1k}+\sum_{k\sim 2,k \nsim 1}w_{2k}\nonumber\\&+&\sum_{k\sim 1,k \sim 2}|w_{1k}-w_{2k}|=\frac{3}{10}
\label{illus5w14b}
\end{eqnarray}
\begin{eqnarray}
W[1,4]&=& w_{1}+ w_{4}+\sum_{k\sim 1,k \nsim 4}w_{1k}+\sum_{k\sim 4,k \nsim 1}w_{4k}\nonumber\\&+&\sum_{k\sim 1,k \sim 4}|w_{1k}-w_{4k}|=\frac{1}{5}
\label{illus5w14a}
\end{eqnarray}
\begin{eqnarray}
W[3,4]&=& w_{3}+ w_{4}+\sum_{k\sim 3,k \nsim 4}w_{3k}+\sum_{k\sim 4,k \nsim 3}w_{4k}\nonumber\\&+&\sum_{k\sim 3,k \sim 4}|w_{3k}-w_{4k}|=\frac{1}{5}
\label{illus5w14c}
\end{eqnarray}
It can be easily shown that for the state $\rho_{5}$, the inequality (\ref{cor-6}) is verified. Thus, we can conclude that the two-qubit state $\rho_{5}$ is a separable state.\\
\textbf{Illustration-5:} Let us consider a $3\otimes 3$ dimensional bipartite state $\rho_{6}$, which is given by
\begin{eqnarray}
\rho_{6}= \frac{1}{N}
\begin{pmatrix}
  x & 0.01 & 0 & 0 & 0 & 0 & 0 & 0 & a \\
  z & x & 0 & 0 & z & 0 & 0 & 0 & 0 \\
 0 & 0 & x & z & 0 & 0 & 0 & 0 & 0 \\
 0 & 0 & z & x & 0 & 0 & 0 & z & 0 \\
 0 & z & 0 & 0 & x & z & 0 & 0 & a \\
 0 & 0 & 0 & 0 & z & x & z & 0 & 0 \\
 0 & 0 & 0 & 0 & 0 & z & y & z & 0 \\
 0 & 0 & 0 & z & 0 & 0 & z & x & 0 \\
 a & 0 & 0 & 0 & a & 0 & 0 & 0 & y \\
\end{pmatrix}
\label{illus6rho}
\end{eqnarray}
where $N=400a+1$, $x=50a$, $y=\frac{50a+1}{2}$, $z=0.01$ and $0.01\leq a\leq 1$.\\
It can be easily checked that the eigenvalues of $\rho_{6}$ are positive for $0.01\leq a\leq 1$ .
The Laplacian corresponding to the density matrix $\rho_{6}$ is given by
\begin{eqnarray}
L_{\rho_{6}}= \frac{1}{N}
\begin{pmatrix}
  u & -z & 0 & 0 & 0 & 0 & 0 & 0 & -a \\
 -z & w & 0 & 0 & -z & 0 & 0 & 0 & 0 \\
 0 & 0 & z & -z & 0 & 0 & 0 & 0 & 0 \\
 0 & 0 & -z & w & 0 & 0 & 0 & -z & 0 \\
 0 & -z & 0 & 0 & v & -z & 0 & 0 & -a \\
 0 & 0 & 0 & 0 & -z & w & -z & 0 & 0 \\
 0 & 0 & 0 & 0 & 0 & -z & w & -z & 0 \\
 0 & 0 & 0 & -z & 0 & 0 & -z & w & 0 \\
 -a & 0 & 0 & 0 & -a & 0 & 0 & 0 & 2a \\
\end{pmatrix}
\label{illus6lap}
\end{eqnarray}
where $u=a+0.01$, $v=a+0.02$ and $w=0.02$.\\
The graph can be constructed from the Laplacian $L_{\rho_{6}}$, which is a connected graph shown in \figref{rho6}. Thus, we calculate $W[1,2], W[1,9], W[2,5],W[3,4], W[4,8], W[5,6]$,
$W[6,7], W[7,8], W[5,9]$ as per the prescription given in (\ref{wij}). Therefore, we have
\begin{eqnarray}
&&W[1,2]=\frac{2a+0.04}{N}, W[1,9]= \frac{4a+0.02}{N},\nonumber\\&&
W[2,5]= \frac{2a+0.06}{N}, W[3,4]=\frac{0.03}{N},\nonumber\\&&
W[4,8]= \frac{0.04}{N}, W[5,6]= \frac{2a+0.06}{N},\nonumber\\&&
W[6,7]=\frac{0.06}{N}, W[7,8]= \frac{0.06}{N},\nonumber\\&&
W[5,9]=\frac{4a+0.04}{N} 
\label{illus6wij}
\end{eqnarray}
The quantity $max_{i \sim j}W[i,j]$ is given by
\begin{eqnarray}
max_{i \sim j}W[i,j]=\frac{4a+0.04}{400a+1},~~ \textrm{when}~~0.01\leq a \leq 1
\label{illus6maxwij}
\end{eqnarray}
\begin{figure}[h]
	\centering
	\includegraphics[scale=0.9]{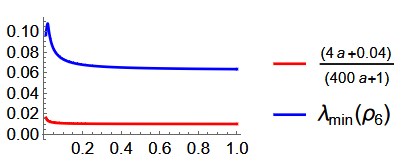}
	\caption{Minimum eigenvalue of $\rho_6$ and $max_{i\sim j}W[i,j]$}
	\label{fig:plot1}
\end{figure}
From \figref{plot1}, it is clear that for the state $\rho_{6}$, the inequality (\ref{cor-6}) is satisfied. Thus, we can conclude that the two-qutrit state $\rho_{6}$ is either a separable state or PPTES i.e. a PPT state.\\
\begin{figure}[h]
	\begin{tikzpicture}
	\centering
	\node [circle, draw](00) at (0,0){1};
	\node [circle, draw](01) at (3,0){2};
	\node [circle, draw](02) at (6,0){3};
	\node [circle, draw](10) at (0,-3){4};
	\node [circle, draw](11) at (3, -3){5};
	\node [circle, draw](12) at (6,-3){6};
	\node [circle, draw](20) at (0,-6){7};
	\node [circle, draw](21) at (3,-6){8};
	\node [circle, draw](22) at (6,-6){9};
	\draw (00) edge [pos=0.5, "$\frac{z}{N}$"] (01);
	\draw (00) edge [bend right, pos=0.5,  "$\frac{a}{N}$"] (22);
	\draw (01) edge [pos=0.25, "$\frac{z}{N}$"] (11);
	\draw (02) edge [pos=0.25,  "$\frac{z}{N}$"] (10);
	\draw (10) edge [pos=0.25,  "$\frac{z}{N}$"] (21);
	\draw (11) edge [pos=0.5,  "$\frac{z}{N}$"] (12);
	\draw (11) edge [pos=0.75,  "$\frac{a}{N}$"] (22);
	\draw (12) edge [pos=0.15,  "$\frac{z}{N}$"] (20);
	\draw (20) edge [pos=0.5,  "$\frac{z}{N}$"] (21);
	
	\end{tikzpicture}
	\caption{Graph corresponding to $L_{\rho_{6}}$}
	\label{fig:rho6}
\end{figure}
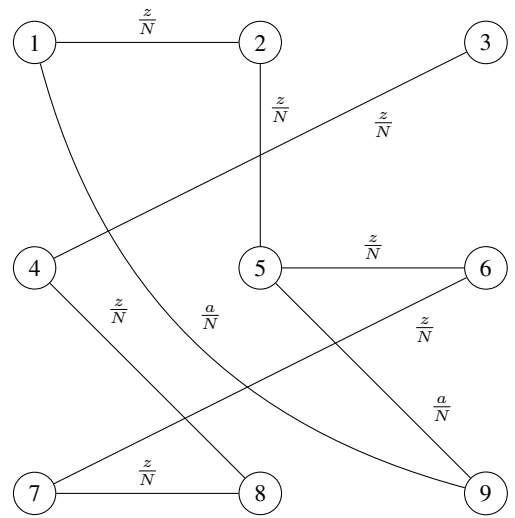

\section{Conclusion}
To summarize, we have constructed a mapping in which the domain set contain any $n\times n$ matrix over the field of real numbers where all off diagonal entries are either positive or negative. In particular, We can consider the domain set as the set of density matrices. The map takes each density matrix into the sum of the input density matrix and another matrix $L$. The matrix $L$ is constructed from the elements of the input density matrix in such a way that it satisfies all the properties of the Laplacian. The constructed map is shown to be unital. Using the unital property and the structure of the map, we are able to derive a criterion that characterize the quantum state either as a pure state or a mixed state. We also use the constructed Laplacian and its partial transpose to derive the PPT criterion. Further, we have derived the inequality between the minimum eigenvalue of the full rank bipartite NPTES and the function $W[i,j]$ of the weights of the edges of a simple connected weighted subgraph of a graph $G$ constructed from the Laplacian. The violation of the derived inequality prove the fact that the given state is a PPT state. Thus, we have studied the entanglement properties of $d_{1} \otimes d_{2}$ dimensional bipartite quantum system through graph theoretical approach. In this work, we restrict ourselves only to the bipartite system but our result is more general in the sense that it can be applied to multipartite system also. Specifically, the result obtained in this work may be useful in studying the three-qubit states in non-inertial frame \cite{qiang1,qiang2,qiang3,qiang4}.
\section{Data Availability Statement}
 Data sharing not applicable to this article as no datasets were generated or analysed during the current study.


\begin{thebibliography}{90}
\bibitem{horodecki1} R. Horodecki, P. Horodecki, M. Horodecki and K. Horodecki, Rev. Mod. Phys. \textbf{81}, 865 (2009).
\bibitem{jozsa} R. Jozsa and N. Linden, Proc. R. Soc. Lond. A \textbf{459}, 2011 (2003).
\bibitem{bennett} C. H. Bennett, G. Brassard, C. Crepeau, R. Jozsa, A. Peres and W. K. Wootters, Phys. Rev. Lett. \textbf{70}, 1895 (1993).
\bibitem{bennett1} C. H. Bennett and S. Wiesner, Phys. Rev. Lett. \textbf{69}, 2881 (1992).
\bibitem{ekert} A. K. Ekert, Phys. Rev. Lett. \textbf{67}, 661 (1991).
\bibitem{ngisin} N. Gisin, G. Ribordy, W. Tittel and H. Zbinden, Rev. Mod. Phys. \textbf{74}, 145 (2002).
\bibitem{peres} A. Peres, Phys. Rev. Lett. \textbf{77}, 1413 (1996).
\bibitem{horodecki2} M. Horodecki, P. Horodecki and R. Horodecki, Phys. Lett. A \textbf{223}, 1 (1996).
\bibitem{rudolph} O. Rudolph, Quantum Inf. Proc. \textbf{4}, 219 (2005).
\bibitem{chen} K. Chen and L.-A. Wu, Quant. Inf. Comp. \textbf{3}, 193 (2003).
\bibitem{horodecki3} M. Horodecki and P. Horodecki, Phys. Rev. A \textbf{59}, 4206 (1999).
\bibitem{guhne} O. Guhne and G. Toth, Physics Reports \textbf{474}, 1 (2009).
\bibitem{braunstein1} S. L. Braunstein, S. Ghosh, T. Mansour, S. Severini and R. C. Wilson, Phys. Rev. A \textbf{73}, 012320 (2006).
\bibitem{braunstein2} S. L. Braunstein, S. Ghosh and S. Severini, Annals of Combinatorics \textbf{10}, 291 (2006).
\bibitem{cabello} A. Cabello, S. Severini and A. Winter, Phys. Rev. Lett. \textbf{112}, 040401 (2014).
\bibitem{ray} M. Ray, N. G. Boddu, K. Bharti, L-C Kwek and A. Cabello, arXiv:2007.10746v2 [quant-ph].
\bibitem{lockhart1} J. Lockhart, O. Guhne and S. Severine, Phys. Rev. A \textbf{97}, 062340 (2018).
\bibitem{dutta} S. Dutta, B. Adhikari, S. Banerjee and R. Srikanth, Phys. Rev. A \textbf{94}, 012306 (2016).
\bibitem{rojo} O. Rojo, Linear Algebra Appl. \textbf{420}, 625 (2007).
\bibitem{anderson} W. N. Anderson and T. D. Morley, Linear and Multilinear Algebra \textbf{18}, 141 (1985).
\bibitem{grone}R. Grone and R. Merris, SIAM J. Discrete Math. \textbf{7}, 221 (1994).
\bibitem{li} J.-S. Li and D. Zhang, Linear Algebra Appl. \textbf{285}, 305 (1998).
\bibitem{merris} R. Merris, Linear Algebra Appl. \textbf{285}, 33 (1998).
\bibitem{pan} Y.-L. Pan, Linear Algebra Appl. \textbf{355}, 287 (2002).
\bibitem{rojo1} O. Rojo, R. Soto and H. Rojo, Linear Algebra Appl. \textbf{312}, 155 (2000).
\bibitem{das1} K. C. Das and R. B. Bapat, Linear Algebra Appl. \textbf{409}, 153 (2005).
\bibitem{das2} K. C. Das, Linear Algebra Appl. \textbf{427}, 55 (2007).
\bibitem{poignard} C. Poignard, T. Pereira and J. P. Pade, SIAM J. Appl. Math. \textbf{78}, 372 (2018).
\bibitem{chung} F. R. K. Chung and R. P. Langlands, J. Combinatorial Th. Series A \textbf{75}, 316 (1996).
\bibitem{lasserre} J. B. Lasserre, IEEE Trans. on Automatic Control \textbf{40}, 1500 (1995).
\bibitem{kumari} A. Kumari and S. Adhikari, Phys. Rev. A \textbf{100}, 052323 (2019).
\bibitem{horn} R. A. Horn and C. R. Johnson, Matrix analysis, (Cambridge University Press, Cambridge, 1999).
\bibitem{kadison} R. V. Kadison, Ann. Math. \textbf{56}, 494-503 (1952).
\bibitem{wolkowicz} H. Wolkowicz and G. P. H. Styan, Lin. Alg. and its Appl. \textbf{29}, 471 (1980).
\bibitem{ekert1} A. K. Ekert, C. M. Alves, D. K. L. Oi, M. Horodecki, P. Horodecki and L. C. Kwek, Phys. Rev. Lett. \textbf{88}, 217901 (2002).
\bibitem{mal} S. Mal, T. Pramanik and A. S. Majumdar, Phys. Rev. A \textbf{87}, 012105 (2013).
\bibitem{phorodecki} P. Horodecki, Phys. Lett. A \textbf{232}, 333 (1997).
\bibitem{nk} M. A. Nielsen and J. Kempe, Phys. Rev. Lett. \textbf{86}, 5184 (2001).
\bibitem{dpd} D. P. DiVincenzo, P. W. Shor, J. A. Smolin, B. M. Terhal and A. V. Thapliyal, Phys. Rev. A \textbf{61}, 062312 (2000).
\bibitem{adhikari} S. Adhikari, Eur. Phys. J. D \textbf{75}, 92 (2021).
\bibitem{qiang1} W-C Qiang, G-H Sun, Q. Dong, and S-H Dong, Phys. Rev. A \textbf{98}, 022320 (2018).
\bibitem{qiang2} Q. Dong, A. J. Torres-Arenas, G-H Sun, W-C Qiang and Shi-Hai Dong , Frotiers of Physics \textbf{14}, 21603 (2019).
\bibitem{qiang3} Q. Dong, A. A. S. Manilla, I. L. Yanez, G-H Sun and S-H Dong, Phys Scr. \textbf{94}, 105101 (2019).
\bibitem{qiang4} Q. Dong ,R. d. J. Leon-Montiel, G-H Sun and S-H Dong, Entropy 24, 1011 (2022). 
\end{thebibliography}
\end{document}